\documentstyle[12pt]{article}

\def\psnormal{\textwidth=16cm\textheight=21.5cm
              \topmargin=0cm\oddsidemargin=0.5cm
              \evensidemargin=0cm\parindent=1cm}
\psnormal
\def\simlt{\stackrel{<}{{}_\sim}}

\def\Eq{\begin{equation}}
\def\End{\end{equation}}

\def\Eqa{\begin{eqnarray}}
\def\Enda{\end{eqnarray}}

\def\Endl#1{\label{#1} \End}

\begin{document}

\renewenvironment{thebibliography}[1]
  { \begin{list}{\arabic{enumi}.}
    {\usecounter{enumi} \setlength{\parsep}{0pt}
     \setlength{\itemsep}{3pt} \settowidth{\labelwidth}{#1.}
     \sloppy
    }}{\end{list}}
\begin{titlepage}

\title{
THE FINE--TUNING PROBLEM OF THE ELECTROWEAK SYMMETRY
BREAKING MECHANISM IN MINIMAL SUSY MODELS\thanks{Talk given at the
XVI Kazimierz Meeting on Elementary Particle Physics: {\em New
Physics with New Experiments}, 24-28 May 1993, Kazimierz (Poland).}}

\author{
B. de CARLOS\thanks{Supported by a Comunidad de Madrid grant.} \\
 Instituto de Estructura de la Materia, CSIC \\
              Serrano 123, 28006-Madrid, Spain
\and
J.A. CASAS\thanks{On leave from Instituto de Estructura de la Materia
(CSIC), Serrano 123, 28006-Madrid, Spain.} \\
Theory Division, CERN \\
CH-1211, Geneva 23, Switzerland}

\date{}

\maketitle

\begin{abstract}
\noindent
We calculate the region of the MSSM parameter space (i.e. $M_{1/2}$,
$m_{0}$, $\mu$, \ldots) compatible with a correct electroweak
breaking and a realistic top-quark mass. To do so we have included
{\em all} the one-loop corrections to the effective potential $V_{1}$
and checked their importance in order to
obtain consistent results. We also consider the fine-tuning problem
due to the enormous dependence of $M_{Z}$ on $h_{t}$ (the top Yukawa
coupling), which is substantially reduced when the one-loop effects
are taken into account. We also explore the reliability of
the so-called "standard" criterion to
estimate the degree of fine-tuning. As a consequence,
we obtain a new set of upper bounds on the MSSM
parameters or, equivalently, on the supersymmetric masses perfectly
consistent with the present experimental bounds.
\end{abstract}

\thispagestyle{empty}
\vspace{-21.0cm}
\begin{flushright} CERN-TH.7024/93\\IEM-FT-78/93\\September 1993\
\end{flushright}
\end{titlepage}

\newpage
\psnormal
\section{Introduction}
\vspace{0.6cm}
The Minimal Supersymmetric Standard Model
(MSSM), is characterized by a Lagrangian
\Eq
{\cal L} = {\cal L}_{SUSY} + {\cal L}_{soft},
\End
where ${\cal L}_{SUSY}$ is the supersymmetric part (derived from
$W_{obs}$, the observable superpotential which includes the usual
Yukawa terms $W_{Y}$ plus a mass coupling between the two Higgs
doublets, $\mu H_{1} H_{2}$) and ${\cal L}_{soft}$ contains the SUSY
breaking terms and is given at the unification scale $M_{X}$ by:
\Eq
{\cal L}_{soft} = - m_{0} \sum_{\alpha} |\phi_{\alpha}|^{2} -
\frac{1}{2} M_{1/2} \sum_{a=1}^{3} \bar{\lambda}_{a} \lambda_{a} -
(Am_{0}W_{Y} + Bm_{0}\mu H_{1} H_{2} + h.c.).
\Endl{soft}
Here $m_{0}$ and $M_{1/2}$ are the universal soft breaking masses
(evaluated at $M_{X}$) for scalars ($\phi_{\alpha}$) and gauginos
($\lambda_{a}$) respectively; $A$ and $B$ stand for the trilinear and
bilinear couplings between scalar fields. So all the
supersymmetric masses are fixed once we have chosen values for the
following MSSM parameters: $m_{0}$, $M_{1/2}$, $\mu$, $A$, $B$,
$h_{t}$, where $h_{t}$ is the top Yukawa coupling (we are neglecting
the influence, small in our case, of the bottom and tau Yukawa couplings).

In particular this set of parameters gives us the form of the Higgs
potential in the MSSM which is responsible for the electroweak
breaking process$^{1}$. By imposing the correct electroweak breaking
scale and a reasonable top-quark mass, the allowed region of values for
these parameters is considerably restricted. Furthermore, if one also
requires the absence of fine-tuning in the value of $h_{t}$ through
the ordinary equation$^{2}$
\Eq
\frac{\delta M_{Z}^{2}}{M_{Z}^{2}} = c \frac{\delta
h_{t}^{2}}{h_{t}^{2}},
\Endl{ft}
by setting an upper bound for $c$, the allowed values for the
parameters are more constrained.

Here we present an analysis of these issues following the recent one
done by Ross and Roberts$^{3}$, but refined with the inclusion
of the one-loop corrections to the effective Higgs potential$^{4}$
(theirs was done considering only the renormalization-improved tree-level
potential $V_{0}$), which gives substantially different results.

\vspace{0.6cm}
\section{Radiative electroweak breaking}
\vspace{0.6cm}
The one-loop Higgs potential of the MSSM, $V_{1}(Q)$, is given at a
scale $Q$ by the sum of two terms: the commonly used
renormalization-improved tree-level potential,
\Eq
V_{0}(Q) = \frac{1}{8} (g^{2}+g'^{2})
(|H_{1}|^{2}-|H_{2}|^{2})^{2} + m_{1}^{2} |H_{1}|^{2} + m_{2}^{2}
|H_{2}|^{2} - m_{3}^{2} (H_{1}H_{2} + h.c.),
\Endl{V0}
with $m_{i}^{2} = m_{H_{i}}^{2} + \mu^{2}$, $i=1,2$
($m_{H_{i}}^{2}(M_{X}) = m_{0}^{2}$) and $m_{3}^{2} = Bm_{0}\mu$;
and the one-loop corrections$^{5}$,
\Eq
\Delta V_{1}(Q) = \frac{1}{64\pi^{2}} Str \left [ {\cal M}^{4} \left
(log \frac{{\cal M}^{2}}{Q^{2}} - \frac{3}{2} \right ) \right ].
\Endl{delta}
Here ${\cal M}^{2}$ is the field dependent tree-level squared mass
matrix which contains {\em all} the states of the theory properly
diagonalised, thus including the dependence on $H_{1}$
and $H_{2}$). All the parameters appearing in these equations are
evaluated at some scale $Q$ and run with it; they can be computed by
solving the standard RGEs$^{6}$, using the present values for the
gauge couplings and taking into account the
supersymmetric thresholds.

To study electroweak breaking we minimize $V_{1}$ to
obtain $v_{1} \equiv \langle H_{1} \rangle$ and $v_{2} \equiv \langle
H_{2} \rangle$. An example of our results can be seen in Fig. 1:
both $V_{0}$ and $V_{1}$ predict electroweak breaking but for
different values of $v_{1}$ and $v_{2}$.
{}From our analysis we see that: {\em i)} the
tree-level approximation
is not reliable, {\em ii)} the top-stop approximation is not accurate
enough to stand for the whole one-loop corrections and {\em iii)}
the complete one-loop solutions are much more stable versus $Q$.

However we may find that, for some values of the parameters, this
stability is

partially spoiled in the region of electroweak breaking ($Q \sim
M_{Z}$) due to large logarithmic corrections. In order to give
general results we choose to take $v_{1}(Q)$ and $v_{2}(Q)$ at some
scale $\hat{Q}$ where $\Delta V_{1}$ is negligible$^{4,7}$ and then
perform the wave function renormalization of the Higgs fields from
$\hat{Q}$ to $M_{Z}$ (which is a small effect indeed).

We can now calculate $M_{Z}$ as
\Eq
(M_{Z}^{phys})^{2} \simeq \left. {\textstyle \frac{1}{2}}
(g^{2}(Q)+g'^{2}(Q))
[v_{1}^{2}(Q) + v_{2}^{2}(Q)] \right|_{Q=M_{Z}^{phys}}
\End
and constrain the MSSM parameters by requiring: {\em a)} correct
electroweak breaking (i.e. $M_{Z}^{phys}=M_{Z}^{exp}$), {\em b)}
reasonable top-quark mass
and {\em c)} absence of electric charge and colour breakdown$^{6}$.
The resulting region of allowed values is enhanced and displaced from
the one obtained in Ref. 3 (see Fig. 2). We have also evaluated the
effect of varying $A$, $B$ and $|\mu/m_{0}|$ with similar
results$^{8,9}$.
\vspace{0.6cm}
\section{The fine-tuning problem}
\vspace{0.6cm}
In the previous section we have restricted the possible values of
the MSSM parameters by imposing a correct electroweak breaking. But
we still can find arbitrarily high values of these parameters
compatible with this constraint, which would lead us to a problem of
fine-tuning$^{3}$. In particular we are interested in the
parameter to which $M_{Z}$ is most sensitive, that is $h_{t}$, and
the degree of fine-tuning is given by Eq. 3 in which $c$ depends on
the whole set of MSSM parameters. So avoiding fine-tuning means
setting an upper bound on $c$, e.g. $c \simlt 10$ as in
Ref. 3 (see Fig. 2a). In our case, the inclusion of the one-loop
corrections soften the dependence of $M_{Z}$ on $h_{t}$, giving for
the same bound, $c \simlt 10$, a broader region of allowed parameters.

But the standard criterion of fine-tuning, Eq. 3, is {\em
ambiguously} defined as it depends on {\em i)} the independent
parameters of the theory and {\em ii)} the physical quantity we are
fitting (note that taking $h_{t}$ ($M_{Z}$) instead of $h_{t}^{2}$
($M_{Z}^{2}$) would change $c$ into $2c$ ($c/2$)). Moreover
we see that it measures sensitivity rather than fine-tuning: a
relationship extremely sensitive between $M_{Z}$

and $h_{t}$ could lead to values of $c$ always higher than the bound,
making this criterion meaningless. We
have checked that fortunately this is not the case in the MSSM$^{8}$.

{}From all these considerations we see that the standard fine-tuning
criterion (3) is more qualitative than quantitative, so we should
conservatively relax the former
bound$^{3}$ up to $c \simlt 20$ at least. This leads us to new upper
limits on the MSSM parameters, as can be seen in Fig. 2b: $m_{0}$, $\mu
\simlt 650$ GeV and $M_{1/2} \simlt 400$ GeV, that imply upper bounds
on the sparticle spectrum:
\begin{eqnarray*}
Gluino    & : & M_{\tilde{g}} \simlt 1100 \ {\rm GeV} \\
Lightest \ chargino & : & M_{\chi^{\pm}} \simlt 250 \ {\rm GeV} \\
Lightest \ neutralino & : & M_{\lambda} \simlt 200 \ {\rm GeV} \\
Squarks & : & m_{\tilde{q}} \simlt 900 \ {\rm GeV} \\
Sleptons & : & m_{\tilde{l}} \simlt 450 \ {\rm GeV}
\end{eqnarray*}

\vspace{0.6cm}
\section*{Figure captions}
\vspace{0.6cm}

{\bf Figure 1:} $v_{1}$ and $v_{2}$ versus the $Q$ scale between
$M_{Z}$ and $2$ TeV (in GeV), for $m_{0}$ = $\mu$ = 100 GeV;
$M_{1/2}$ = 180 GeV; $A$ = $B$ = 0; $h_{t}$ = 0.250. Solid lines:
complete one-loop results; dashed lines: "improved" tree-level
results; dotted lines: one-loop results in the top-stop approximation.

\noindent
{\bf Figure 2:} The case $A = B$ = 0, $|\mu/m_{0}|$ = 1 with (a) the
tree-level potential $V_{0}$ and (b) the whole one-loop
effective potential $V_{1}$. Diagonal lines correspond to the extreme
values of $m_{t}$ as were calculated in Ref. 3: $m_{t}$ =
160, 100 GeV. Transverse lines indicate constant values of $c$.

\end{document}